\documentclass[doublecol]{epl2} 
\usepackage{amsmath}
\def\sgn{\mbox{sgn}}

\title{ Memoryless nonlinear response: A simple mechanism for the $1/f$ noise}

\author{Avinash Chand Yadav\inst{1} \and Ramakrishna Ramaswamy\inst{1,2} \and Deepak Dhar\inst{3}}
\shortauthor{Yadav \etal}

\institute{                    
  \inst{1} School of Physical Sciences, Jawaharlal Nehru University, New Delhi 110 067, India\\
  \inst{2} University of Hyderabad, Hyderabad 500 046, India\\
  \inst{3} Department of Theoretical Physics, Tata Institute of Fundamental Research,\\ Homi Bhabha Road, Mumbai 400 005, India
}

\abstract{Discovering the mechanism underlying the ubiquity of $``1/f^{\alpha}"$ noise has been a long--standing problem. The wide range of systems in which the fluctuations show the implied long--time correlations suggests the existence of some simple and general mechanism that is independent of the details of any specific system. We argue here that a {\it memoryless nonlinear response} suffices to explain the observed non--trivial values of $\alpha$: a random input noisy signal $S(t)$ with a power spectrum varying as $1/f^{\alpha'}$, when fed to an element with such a response function $R$ gives an output $R(S(t))$ that can have a power spectrum $1/f^{\alpha}$ with $\alpha < \alpha'$. As an illustrative example, we show that an input Brownian noise ($\alpha'=2$) acting on a device with a sigmoidal response function $R(S)= \sgn(S)|S|^x$, with $x<1$, produces an output with $\alpha = 3/2 +x$, for $0 \leq x \leq 1/2$. Our discussion is easily extended to more general types of input noise as well as more general response functions.}

\begin{document}

\maketitle

Although first observed almost ninety years ago and subsequently found to occur in very diverse systems, the origins of the ubiquitous low frequency noise with power spectrum $1/f^{\alpha}$ where $\alpha \approx 1$--- called flicker noise or pink noise--- has been a long--standing conundrum \cite{Dutta_1981}. Some examples are the fluctuations in voltage across a resistor or other components in electronic equipment \cite{Dutta_1981}, river discharges \cite{Wang_2008, Thompson_2012}, traffic flow \cite{Musha_1976}, the frictional force in sliding friction under wear conditions \cite{Durate_2009}, and the acoustic power in music or speech \cite{Voss_1975}. Many signals involving response to physical stimuli in living systems, such as fluctuations in the response time of motor-response to a periodic stimulus in humans \cite{Chen_1997}, in the time series of errors of replication of spatial or temporal intervals by memory in humans \cite{Gilden_1995}, or in human colour vision \cite{Medina_2012}  have also been found to have $1/f^{\alpha}$ spectrum. While it seems unlikely that these very diverse types of processes could share a single underlying mechanism, it also seems reasonable that the number of mathematical mechanisms underlying this behaviour would not be very large. Considerable effort has therefore been devoted to discovering these \cite{Dutta_1981}.

In this Letter, we point out that {\it memoryless nonlinear response} (MNR) is sufficient to explain the observed nontrivial values of $\alpha$. Despite its simplicity, this very general mechanism does not appear to have been explicitly discussed in the literature so far. When the response--time of the system is much shorter than the time--scale of fluctuations under discussion, we may model the system as a nonlinear device which when subjected to the input signal produces a response that is a nonlinear function of the instantaneous value of the input noise. Thus if $S(t)$ is some noise process, we obtain another process $R(t)$, whose value at any time $t$ is related to the value of $S(t)$ at the same time by the transformation: $R(t)= h(S(t))$, where $h$  is a simple nonlinear function. The process $R(t)$ can have a different value of the spectral power exponent, and we expect that this gives rise to the observed nontrivial values of $\alpha$ in many situations. 

\begin{figure}
\centering
\scalebox{0.65}{\includegraphics{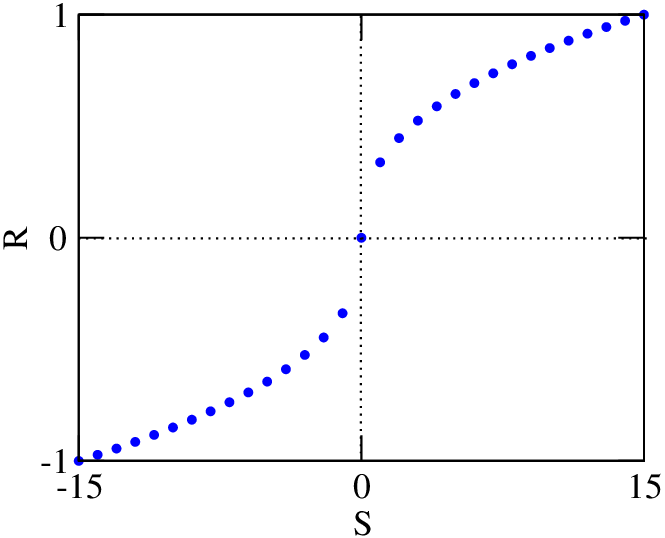}}
\caption{A typical response versus stimulus curve for the nonlinear function in eq.~(\ref{eq1}) with nonlinearity parameter $x=0.4$.}
\label{fig1}
\end{figure}  

The importance of MNR has not been discussed in the literature on $1/f^{\alpha}$ noise, perhaps partly due to the fact that in the prototypical instances such as for voltage fluctuations in resistors, the device is being operated in a clearly linear response regime, and nonlinear effects are expected to be unimportant. This is not so in other applications such as those mentioned above. 

The example we consider below is the response of a non--linear  system driven by an input noisy signal. We assume that input signal is a bounded  Brownian noise, having a $1/f^2$ spectrum.  When this acts on a device with a sigmoidal response function, $R(S)= \sgn(S)|S|^x$, the resulting output process has the spectral exponent $\alpha = 3/2 +x$, for $0\leq x \leq 1/2$.

It is useful to contrast this with other explanations of $1/f$ noise that have commonly been proposed earlier, and which can be grouped into three broad classes. The best--known theoretical description treats the noisy signal as a superposition of many small burst--like events such as ``shot noise'' in resistors and thermionic valves or the record of seismic activity in an area. In the simplest discussions, these pulse--like events are taken to be independent, and with additional assumptions about distribution of shapes and amplitudes of these pulses, one can get an $1/f^{\alpha}$ spectrum \cite{Dutta_1981}. To get a broad distribution of relaxation times, specific mechanisms such as the relaxation time being a product of several random variables and hence distributed log-normally have been proposed \cite{Montroll_1982}.  

In the second approach, one models the system in which transport occurs as an extended driven system with some reasonable evolution rules. Diffusive transport may be modeled by stochastic nonlinear partial differential equations, for instance, and Grinstein {\it et al.} \cite{Grinstein_1992} found that in a model of particle transport in contact with reservoirs, fluctuations in the total number of particles show a $1/f$ or $1/f^{3/2}$ power spectrum, depending on whether the system was driven only from the boundary or in the bulk. Discrete models of nonlinear transport such as the well studied Abelian sandpile automaton \cite{Bak_1987, Dhar_2006}, when driven by random addition of particles give rise to a critical steady state in which the fluctuations in the total activity or the mass of the pile show $1/f^{\alpha}$ fluctuations \cite{Laurson_2005, Davidsen_2002}. The power spectrum of activity in such models can be determined in terms of the distribution of inter-pulse intervals and the detailed characteristics of the pulse shape. Mass fluctuations in a one--dimensional sandpile have been shown to give a $1/f$ power spectrum over a very wide range of $f$ \cite{Maslov_1999, Yadav_2012}. 

\begin{figure}
\centering
\scalebox{0.55}{\includegraphics{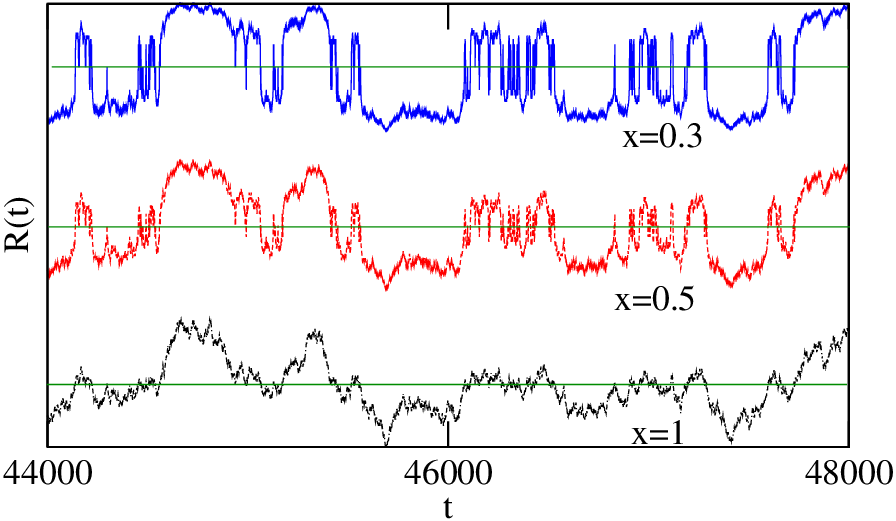}}
\caption{Signals obtained by applying nonlinear transformations to a bounded Brownian signal for different values of the parameter $x$. The value $x=1$ corresponds to the identity transformation, namely the unmodified input signal. A segment of the noisy time series is shown.}
\label{ts}
\end{figure}

The third class identifies a small number relevant degrees of freedom, and writes down their evolution equations, usually coupled stochastic differential equations. Signals deriving from chaotic dynamical systems can yield $1/f^{\alpha}$ even with a small number of degrees of freedom \cite{Benmizrachi_1985}. Nonlinear flows subject to multiplicative noise can also generate a $1/f^{\alpha}$ spectrum with the exponent being in the range $0.5 < \alpha < 2$ \cite{Ruseckas_2010}. There have been proposals to treat $1/f^{\alpha}$ noise as a generalized renewal process \cite{Lowen_1993, Kaulakys_2005}. 

The case we treat here is that of an input Brownian noise, although our analysis is valid for considerably more general cases. Since unbounded Brownian motion is not stationary, we take $S(t)$ to be a bounded random walk. To control the singularities of the Brownian walk at short length scales, we impose a lattice cutoff: the walker moves on a linear chain of $2B + 1$ points and takes steps $\pm 1$ with equal probability at each time step $t$. At the end--points of the interval $\pm B$, the walker is reflected with unit probability. 

The response is often sigmoidal in shape. Two instances from psychology are representative \cite{Copelli_2002}: the phenomenological Stevens law states that the response is a nonlinear function of the stimulus of the power--law type, and the Weber--Fechner law posits a logarithmic function, namely $R(S) \sim \log (S)$ \cite{Stevens_1975}. More generally, when the strength of the stimulus can vary over several orders of magnitude, the response would be expected to show lower sensitivity at larger signal strengths. We note that receptors in the eye, for instance, can respond to light levels that vary over eight orders of magnitude \cite{Rose_1948}, and the receptors involved in bacterial chemotaxis can function when the concentration of attractants varies over about six orders of magnitude \cite{Lamanna_2002}. 

We consider the simple functional form shown in fig.~\ref{fig1},
\begin{equation}
R(S)  = \sgn(S)|S|^x,
\label{eq1}
\end{equation}
where $x$ is a parameter $0 \leq x \leq 1$. Other functional forms that have this power--law behaviour for $1 \ll |S| \ll B$ would behave similarly. Note that $S(t)$ takes only integer values, and the apparent singularity of the functional form in eq.~(\ref{eq1}) at $S=0$ is immaterial. A typical signal and its response for different values of $x$ is shown in fig.~\ref{ts}. 
  
In steady state the autocorrelation function $C(\tau)$ for the response $R$ is given by 
\begin{equation}
C(\tau) = \langle R(t)R(t+\tau)\rangle/ \langle R^2(t)\rangle,
\end{equation}
where the angular brackets $\langle \cdot \rangle$ denote average in the steady state.

\begin{figure}
  \centering
  \scalebox{0.65}{\includegraphics{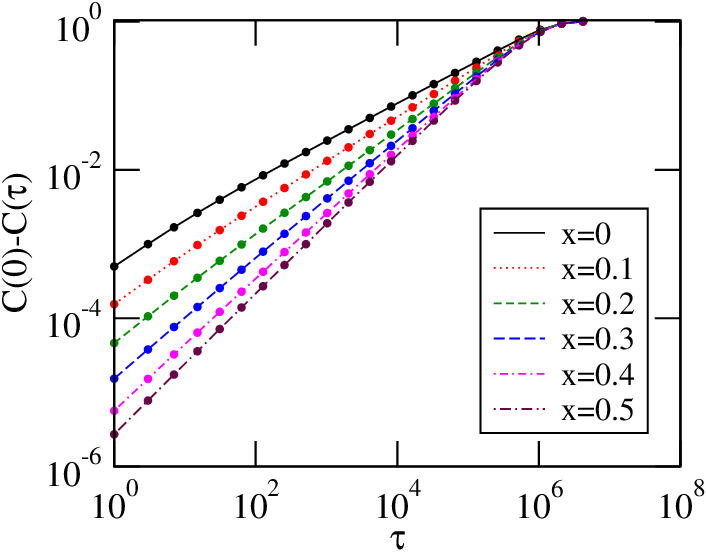}}
  \caption{The correlation function for different values of $x$. For time--series of length $10^8$, the ensemble average is taken over $10^2$ samples. We take $B=2^{10}$, and $\tau$ is chosen in steps of $2^i$, where $i$ runs from 1 to 23.}
\label{cr}
\end{figure}

\begin{figure}
  \centering
  \scalebox{0.65}{\includegraphics{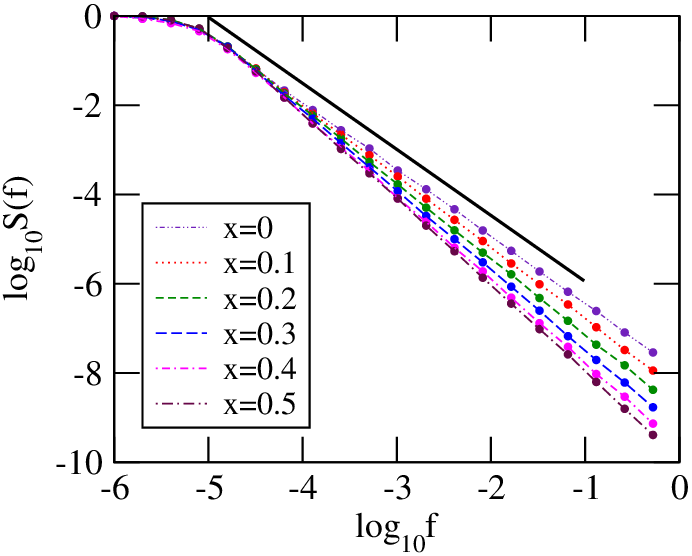}}
  \caption{Power spectra for different values of $x$. Here $B=2^9$ and each point is obtained by ensemble averaging over $10^3$ signals. The reference line has slope 3/2.}
\label{ps}
\end{figure}

For $x$ = 0, the signal only takes values $\pm 1$ or 0, and determining $C(\tau)$ is rather elementary. We calculate the conditional expectation value $\langle \sgn[z(t)]\rangle_{z_0}$ and then average this over values of $z_0$. Since for a walker that crosses the origin at any time, the future evolution is equally likely to be on either side, this reduces to calculating the probability that the walker does not cross $z$ = 0 for the next $\tau $ steps. This is a well-known problem \cite{Feller_1971, Kampen_2001}, and gives   
\begin{equation}
C(\tau) = 1 -\frac{\sqrt{\tau}}{B} F(\tau/B^2)~~~ \text{for $x$ = 0},
\end{equation} 
for $1\ll\tau\ll B^2$, where the scaling function $F(\xi)\to$ constant for $\xi\ll 1$. The $\sqrt{\tau}$ singularity in $C(\tau)$ shows up as $1/f^{3/2}$ dependence for $1 \gg f \gg 1/B^2$. 

For the case $x>0$, $C(\tau)$ can be calculated using the method of images \cite{Feller_1971, Kampen_2001}. The problem reduces to a random walk with an absorbing boundary at $z=0$  and a reflecting boundary at $z=B$. If we start with a source particle $+q$ at $z_1$, the image source $-q$ is at $z'=-z_1$ and $+q$ is at $z'' = 2B-z_1$. These images in turn have their images at $z''' = 2B+z_1$ of strength $-q$, and at $z'''' = -2B+z_1$ of strength $-q$, and so on. 

The probability $P(z_2,\tau|z_1,0)$ of the random walker being at $z_2$ after $\tau$ time steps, given that the initial position is $z_1$ is calculated by the sum of image solutions as
\begin{eqnarray}
\nonumber
P(z_2,\tau|z_1,0) = \sum_{\sigma =\pm 1} \sum_{ r = -\infty}^{+\infty} [\sigma G_0(z_2- \sigma z_1 - 4rB,\tau)\\
+ \sigma G_0(z_2+ \sigma z_1 - 2B-4rB,\tau)].
\end{eqnarray}
The Gaussian propagator $G_0$ is given by
\begin{equation}
G_{0}(z_2,\tau|z_1,0) = \frac{1}{\sqrt{2\pi\tau}}\exp\left[-\frac{(z_2-z_1)^2}{2\tau}\right].
\end{equation}
In calculating the expectation value of $\langle z_{1}^{x} z_{2}^{x}\rangle$, the contribution from the first term is 
\begin{equation}
\frac{1}{\sqrt{2\pi\tau}}\int_{0}^{B} dz_1 \int_{0}^{B} dz_2   z_1^x z_2^x \exp\left[-\frac{(z_2 - z_1-4rB)^2}{2\tau}\right].\nonumber
\end{equation}
Upon changing the variables $y=z/\sqrt{2\tau}$ and using scaling arguments, this gives the leading behaviour of the $r=0$ term as $\sim \tau^{\frac{1}{2}+x}F_1(\tau/B^2)$, where $F_1(\cdot)$ is a scaling function. This straight forward calculation shows that for $1 \ll \tau \ll B^2$ the leading dependence of $C(\tau)$ on $\tau$ is given by 
\begin{equation}
C(0)-C(\tau) \sim \tau^{\frac{1}{2}+x}F_1(\tau/B^2)+ K \tau + {\rm higher~ order~ terms},
\end{equation}
where $K$ is some constant, and $F_1(\xi)$ is a scaling function, which tends to a finite constant as $\xi$ tends to zero. This small $\tau$ behaviour of $C(\tau)$ leads to the power spectrum behaving as $1/f^{\alpha}$ for $1/B^2 \ll f \ll 1$, with 
\begin{equation}
\alpha = 
\begin{cases} 3/2 + x,& {\rm ~for~} 0\leq x \leq 1/2\\
         2,&  {\rm ~for~} 1/2 \leq x \leq 1.
\end{cases}
\end{equation}

We have verified the above analysis numerically through Monte Carlo simulations. A realization of $S(t)$ is generated from the trajectory of a random walker with reflecting boundaries (a portion of this is shown in fig.~\ref{ts}) and this is transformed to the time--series $R(t)$ whose power spectrum is then determined using standard fast--Fourier transform methods. Figure~\ref{cr} shows the autocorrelation function of the response signal for different values of $x$. The variation of the exponent $\alpha$ with $x$ is shown in fig.~\ref{ps}, for different values of $x$, using $B = 2^9$. We see that the effective exponent $\alpha$ is nearly constant for the entire range $1/B^2 \ll f \ll 1$ (approximately five decades of frequency), and the slopes are in good agreement with the analytical prediction.    

Clearly, $S(t)$ satisfies the linear equation 
\begin{equation}
\frac{d}{dt} S(t) = \eta(t),
\end{equation}
where $\eta(t)$ is a white noise. Expressing $S(t)$ in terms to $R(t)$, it is easy to see that $R(t)$ satisfies a stochastic differential equation with a {\it multiplicative nonlinear white noise}. Nonlinear differential equations with multiplicative white noise have been studied earlier \cite{Ruseckas_2010}, and are known to give $1/f^{\alpha}$ type noise. The present treatment is, however, more general. 

\begin{figure}
  \centering
  \scalebox{0.60}{\includegraphics[angle=0]{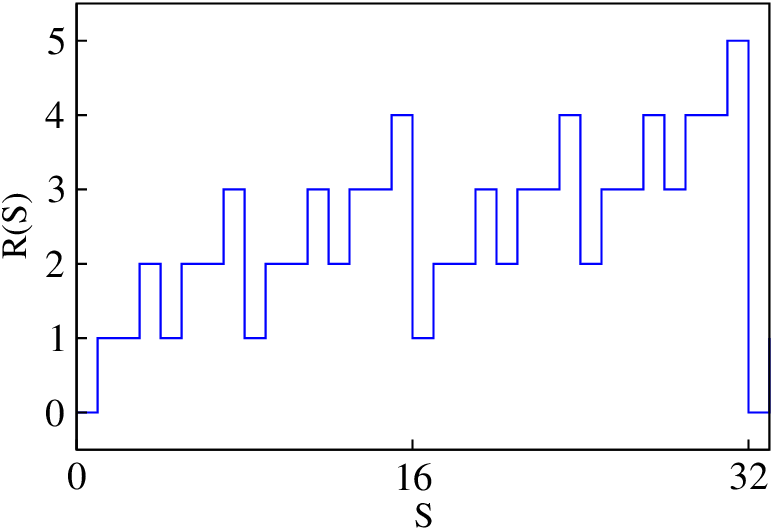}}
  \caption{A more complicated nonlinear response function that gives $1/f$ noise: the signal $S$ is discretized to 32 levels, and $R(S)$ is the number of 1's in the binary representation of the discretized $S$.}
\label{fig5}
\end{figure}

Our analysis applies for both a wider class of input and a wider class of response. While the sigmoidal shape of the nonlinear transformation $R(S)$ shown in fig.~\ref{fig1} is encountered most often, it is not a necessary condition for the mechanism discussed here. If the response depends on some specific feature of the input, its description may require a more general function $R(S)$ than the form chosen in eq.~(\ref{eq1}). We take $R(S)$ for $S \in (-B,+B)$ to be defined by its Fourier expansion on the range $[-B, 3B]$ (making it symmetric about $S=+B$ to avoid a discontinuity) 

\begin{equation}
R(S) = \sum_{n=-\infty}^{+\infty} r_n \exp[ i \pi n S/(2 B)].
\label{ftrs}
\end{equation}
Note that this Fourier series is in the {\it amplitude} of the signal, and not directly in its time dependence. 

Let $g(a,\tau) = \langle \exp \left[ i a (S(t) -S(t +\tau))\right] \rangle $. Then, it is easy to see that
\begin{equation}
C(\tau) = \sum_{n} |r_n|^2 g\left(\frac{n \pi}{2 B}, \tau\right).
\end{equation}
By using the fact that for the bounded Brownian signal $S(t)$, $g(a, \tau) \approx \exp \left[ - a^2 |\tau|/2 \right]$, for $\tau \ll B^2$, we get  
\begin{equation}
C(\tau) \approx  \frac{1}{\langle R^2(t) \rangle }\sum_{n=-\infty}^{+\infty} |r_n|^2 \exp[ - \pi^2 n^2 |\tau|/4 B^2].
\label{crg}
\end{equation}
Taking Fourier transforms, we get
\begin{equation}
P(f) \sim  \frac{1}{\langle R^2(t) \rangle }\sum_{n=-\infty}^{+\infty}
\frac{n^2 B^2 |r_n|^2}{ \pi^4 n^4 + 16 f^2 B^4}.
\end{equation}

The function shown in fig.~\ref{fig5} is a response that is a sum of square waves of periods 2, 4, 8, 16 and 32. Such a functional form arises in the study of toppling activity in an Abelian sandpile on a strip, and measures the number of zeros in a binary representation of a random walk on a ring \cite{Yadav_2012}. For this $R$, it can be shown that the power spectrum varies exactly as $1/f$ for $1/B^2 \ll f \ll 1$. For different functional dependence of $r_n$ on $n$, we can get different power--law exponents $\alpha$ for $1/B^2 \ll f \ll 1$, and indeed, the entire range of values $1 \le \alpha \le 2$ may be attained.

The above argument can be generalized to other cases. For a non--Brownian input signal, only the functional form $g(a, \tau)$ is changed, but by varying the dependence of $|r_n|^2$ on $n$, we can get different power--law exponents $\alpha$. The discussion is also easily extended to the case where the observed signal $R(t)$ depends in a nontrivial way on more than one input signals. Say $R(t)$ is a
nonlinear function of two signals $S_1(t)$ and $S_2(t )$ so that we may write $R(t) = R(S_1(t), S_2(t - \Delta))$, where $\Delta $ is some fixed time-delay. Consequently, in contrast to eq.~(\ref{ftrs}), $R(t)$ is a double Fourier series
\begin{equation}
R(S_1,S_2) =\sum_{m,n} r_{m,n} \exp[ i \pi ( m S_1 + n S_2)],
\end{equation}
and then the correlation function for $R$ [cf. eq.~(\ref{crg})] will become a double sum over indices $m$ and $n$ of a similar function involving the correlations of $S_1(t)$ and $S_2(t)$. The behaviour of power spectrum of $R(t)$ is then determined by the
properties of the functions $|r_{m,n}|^2$, and $g_{i,j}(a, \tau)$ (with $i,j$ = 1, 2).

To summarize, understanding the ubiquity of $1/f$ noise has been a problem of long--standing interest. A number of mechanisms have been proposed in the past, but a plausible and general explanation for the wide range of systems has been lacking. Our proposal here is a simple one: in many cases, the nontrivial values of $\alpha$ in the observed $1/f^{\alpha}$ spectrum can be understood as a consequence of the nonlinearity of the response of the system to an input noisy signal.

\acknowledgments
ACY would like to acknowledge CSIR, India for financial support. RR and DD have been partially supported by the DST, India through the J C Bose Fellowship. DD thanks NICK-JONES  and G. S. AGARWAL for some discussions on this topic, and KEDAR DAMLE and J. K. JAIN for critical comments on the manuscript.

\end{document}